# Lightweight Carbon Fiber Mirrors for Solar Concentrator Applications

Nina Vaidya[1], Michael D. Kelzenberg[1], Pilar Espinet-Gonzalez[1], Tatiana G. Vinogradova[1,2], Jing-Shun Huang[1], Christophe Leclerc[1], Ali Naqavi[1], Emily C. Warmann[1], Sergio Pellegrino[1], Harry A. Atwater[1]

[1]California Institute of Technology, Pasadena, CA 91125, United States
[2]Northrop Grumman Aerospace Systems, Azusa, CA 91702, United States

*Abstract* — Lightweight parabolic mirrors for solar concentrators have been fabricated using carbon fiber reinforced polymer (CFRP) and a nanometer scale optical surface smoothing technique. The smoothing technique improved the surface roughness of the CFRP surface from ~3 μm root mean square (RMS) for as-cast to ~5 nm RMS after smoothing. The surfaces were then coated with metal, which retained the sub-wavelength surface roughness, to produce a high-quality specular reflector. The mirrors were tested in an 11x geometrical concentrator configuration and achieved an optical efficiency of 78% under an AM0 solar simulator. With further development, lightweight CFRP mirrors will enable dramatic improvements in the specific power, power per unit mass, achievable for concentrated photovoltaics in space.

## I. Introduction

We are seeking to develop technologies that enable cost-effective space-based solar power (SSP). SSP has long been proposed to meet earth's baseload electrical power needs with solar energy, by operating large-scale solar power stations in space and beaming the energy wirelessly to earth [1] or utilizing the DC power in space. The building block of our proposed power station is the 'tile,' depicted in Figure 1: a ~10 x 10 cm modular element which performs solar photovoltaic energy collection, conversion to radio frequency energy, and transmission of the energy towards earth-based receivers.

Due to high space launch costs, the key challenge is to increase the specific power, or power per unit mass, of SSP technology. Lightweight concentrating optics can increase the specific power of space photovoltaic (PV) energy converters many fold, because reflective or refractive optics can generally be realized at dramatically lower area density (mass per unit area) than can solar cells and their radiation shielding [2]. Parabolic mirrors have been used extensively for concentrated photovoltaics, including in space applications [3, 4]. Here, we report a lightweight parabolic mirror fabrication process based on using cast carbon fiber reinforced polymer (CFRP) parabolas with a surface smoothing technique [5] to produce specular and highly reflective mirror surfaces.

## II. Design & fabrication

The PV concentrator shown in Figure 1 comprises parallel parabolic mirror troughs (parabolic in one plane and linear along the length) each having a focal line at the top back edge of the neighboring mirror. A row of multi-junction solar cells is attached at this point to collect the focused sunlight. This geometry is particularly well suited for space applications because the mirror troughs are foldable for efficient packing prior to launch, and also provide heatsinking, radiative cooling, and radiation shielding for the cells [4, 6].

This paper focuses on the fabrication and testing of the parabolic reflectors for this tile concentrator concept. We have investigated two fabrication approaches for the parabolic reflectors. Initially, we used metalized Kapton membranes as the reflectors, with parabolic supports placed at either end of the trough to impart the correct shape. Kapton is a well-known space-grade polymer and is commercially available in thin sheets with relatively smooth and specular surfaces. However, it was difficult to fabricate the relatively thick metal layer (2–10 μm Al), which is required for thermal conductivity, without degrading the specular reflectance and shape accuracy of the Kapton membrane reflectors. Furthermore, once the cells were mounted at the back edge of the kapton parabolic troughs, the shape deformed from the attachment and thermal stressed of the cells.

To improve the shape accuracy and optical efficiency, we used thin CFRP to fabricate the reflectors. Carbon fiber composites have excellent strength to mass ratio and find much use in aerospace. For our application, thin CFRP sheets are particularly promising because (a) they can be cast to the desired shape, (b) they are flexible but spring back to shape, (c) bare CFRP typically has high thermal emissivity, and by correct choice of fiber type and orientation, they can offer high in-plane thermal conductivity [7]. However, thin CFRP castings are difficult to form into precision shapes for optical applications, and typically have rough and non-specular surfaces which are unsuitable for direct use as a mirror substrate [8].

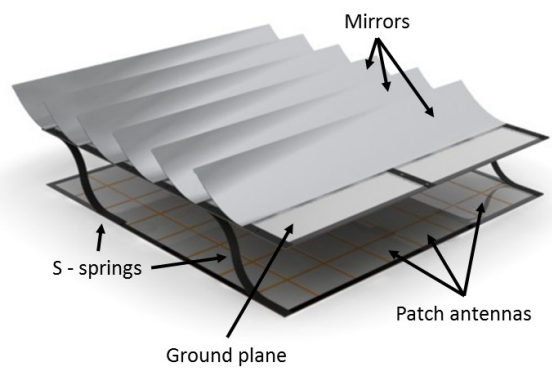

Fig. 1. Conceptual rendering of space solar power tile.

The composite parabolic reflectors were manufactured from unidirectional tape (T800 prepreg, 17 g/m$^2$), arranged in 8 plies with stacking sequence [0/90/+45/−45]$_S$. A steel mold was machined, providing a convex surface of the desired parabolic profile extended by tangential flats on both sides. After lay-up, the composite was vacuum-bagged and cured in an autoclave furnace. Due to the elevated temperatures involved, thin composite materials tended to deform after curing because of imbalanced thermal stresses. Thus an iterative process was used to create a mold which yielded the desired parabolic profile in the castings. The shape of each casting was measured with a FARO ScanArm 3D scanner tool, and if necessary, another mold was machined to correct for any systematic shape errors observed.

Although this process produced CFRP sheets of the desired shape, their surfaces were rough, and they exhibited a diffuse optical appearance. The next challenge was to create a high-quality specular mirror on the surface of the CFRP castings, without deforming the parabolic shape. The surface roughness issue was solved by a novel smoothing technique [5], in which a resin mixture is applied to the surface and allowed to settle. Surface tension produces a smooth and conformal surface. The ultraviolet (UV) cure process minimizes shrinkage of the polymer, which maximizes smoothness and shape accuracy. A completed mirror is shown in Figure 2(a).

The procedure for creating mirrors from CFRP castings was:
1. Clean part, then oven dry.
2. Apply a thin layer of polymer on the surface.
3. Degas in vacuum chamber.
4. Brush off excess polymer and degas again, if necessary.
5. Cure with UV exposure.
6. Deposit reflector layers onto the UV-cured surface

The reflector layers applied to the smoothing polymer comprised a 10 nm Cr adhesion layer, a 120 nm Ag reflector layer, and a 10 nm SiO$_2$ protective layer. All were deposited by electron beam evaporation. Prior to smoothing or metallization, the CFRP average thickness was about 180 μm.

### III. SHAPE CHARACTERIZATION

To determine the shape accuracy and performance potential of the CFRP reflectors, we scanned their shape using a FARO ScanArm 3D scanner. This produced point cloud data describing the mirror surfaces with ~25 μm accuracy, which were recorded with a point density of ~2500/cm$^2$. A typical point cloud data set for the fabricated mirror is plotted in Figure 2(b).

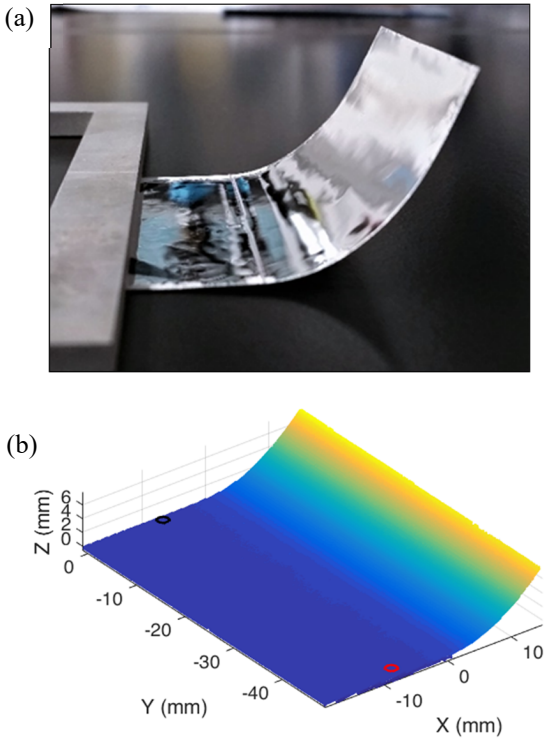

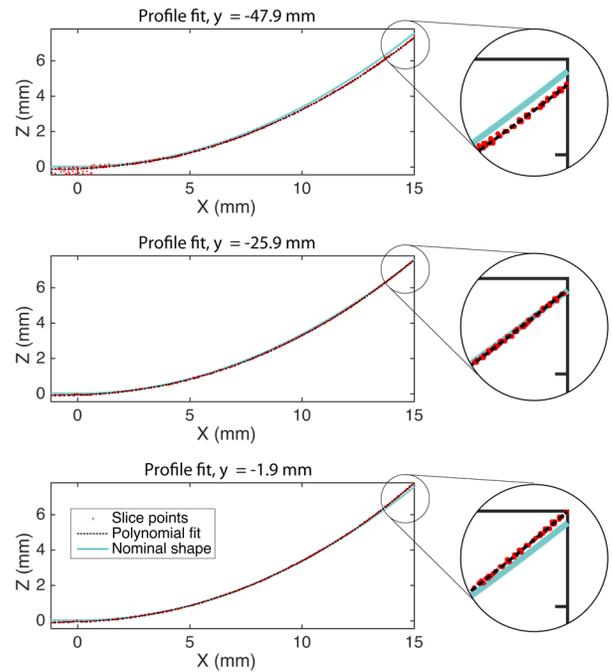

Fig. 2. (a) Photograph of the fabricated CRFP mirror after smoothing and Ag deposition. (b) Point cloud data for the same specimen acquired using a 3D scanner. The individual points are not distinguishable at this resolution. The coordinate system is aligned such that the nominal vertex of the parabolic profile occurs at $x = 0$ and $z = 0$. The raw data was cropped at $x = 15$ mm. The colormap is indexed to the $z$-value of each point. The black and red circles indicate the position of the left and right reference features, respectively, which were used to define the measurement coordinate system i.e., same as the raytracing coordinate system.

Fig. 3. Polynomial fit to point cloud data, and comparison to nominal parabolic profile, for three selected slices within the point cloud data set: the rightmost slice (top), the center slice (middle), and the leftmost slice (bottom). Plots use coordinate system of Figure 2(b).

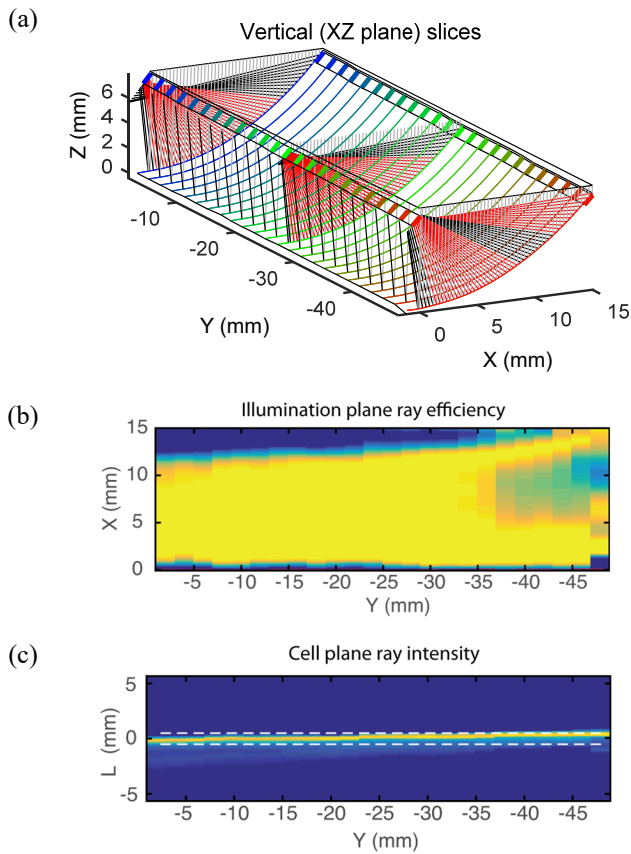

Fig. 4. Ray tracing analysis for 2° incidence angle. All plots use the coordinate system of Figures 2(b) and 3.
(a) 3D plot showing the shape ('slices') used for ray tracing. The receiver cell positions (1 mm width) are also shown. A *y*-indexed color gradient is applied to better illustrate depth. Ray tracing paths are illustrated for array of rays the leftmost, center, and rightmost analysis planes, and for just the first ray of each slice (that is, nearest *x*=0). Incident rays are **gray**, reflected rays which strike the receiver cell are **red**, and reflected rays which miss the receiver cell are **black**.
(b) Illumination plane ray efficiency plot, indicating which areas of the reflector successfully reflect incident light to the receiver cell (yellow), and which areas of the reflector cause the light to miss the receiver cell (blue). Intermediate values occur due to the finite angular width of the source considered (1.5° disk[*]).
(c) Cell plane ray intensity plot, showing the intensity of light reaching the cell plane. The extent of the cell (1 mm) is indicated by the white dashed lines at $L = \pm 0.5$ mm.

To evaluate the accuracy of the mirror shape, we compared cross-sections of the surface data to the desired (nominal) parabolic profile. To assess the impact of the shape on the mirror's utility as a PV concentrator, we performed 2.5D raytracing to calculate the potential optical efficiency.

Following coordinate system alignment, the point cloud data were split and flattened into 2D *x-z* cross sections ('slices')

[*] We use a source width i.e., angular radius of 1.5° because it best approximates the illumination from our solar simulator. The sun's angular radius is ~0.25°.

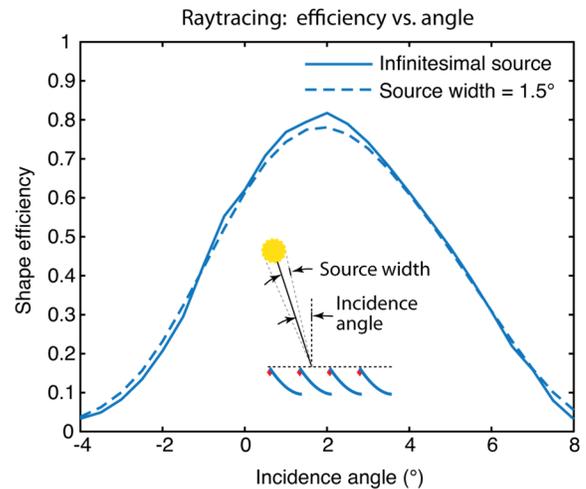

Fig. 5. Angular acceptance versus shape efficiency calculated by ray tracing.

along the length of the reflector. In the illustrated case, the slice width was 2 mm. Then, high-order polynomials were fit to the 2D point data for each slice (8th order fits were used). Figure 3 shows the slice point data, the polynomial fit, and the nominal parabolic profile, for three selected slices within the data set: the rightmost slice (top), the center slice (middle), and the leftmost slice (bottom).

Figure 4(a) shows the polynomial fits and receiver cell position for all slices in the data set. The data is displayed in 3D coordinates by plotting each 2D profile at the *y*-value corresponding to that slice's center plane. The receiver cell positions were determined by calculating where cells would be mounted on the back side of the measured reflector, then the cell positions were translated by a fixed *x* offset corresponding to the reflector pitch in the concentrator design (here, 15 mm) which positioned the cells at the focal line of the reflector.

Shown in Figure 4(a) is a down-sampled ray diagram for each of three slices featured in Figure 3. Red colored rays reach the receiver cell, while black colored rays miss. Figure 4(b) shows the ray efficiency map for all slices as an intensity plot. The correspondence of the left, center, and right-most columns of the image in Figure 4(b), to the ray diagrams above in Figure 4(a), is apparent.

Figure 4(c) shows the intensity distribution of rays reaching the cell plane, relative to the centerline of the cells in said plane (which defines *L*=0 in this image). It is observed that the position of peak intensity shifts slightly, relative to the cell position, over the length of the concentrator. The cause of this shift is evident from examining the three shape cross-sections plotted in Figure 3. A slight twist in the reflector shape has caused the focal line to become misaligned with the nominal focal line i.e., cell centerline position.

For this CFRP mirror, at an incidence angle of 2°, raytracing predicts that ~80% of incident light upon specular reflection from the mirror should reach the receiver cell (see Figure 5). We call this value the *shape efficiency* to distinguish it from true optical efficiency. Note that shape efficiency does not include scattering and absorption losses in the reflector, skewed rays, nor any shading or reflectance losses at the cell; furthermore, several simplifying assumptions have been made in its calculation. Nevertheless, shape measurements and shape efficiency calculations have proven to be useful analysis techniques in our pursuit of improved concentrator design and performance.

## IV. SURFACE ROUGHNESS CHARACTERIZATION

An as-cast CFRP surface was characterized using laser scanning confocal microscopy due to the scale of the roughness of the CF surface. The RMS surface roughness was found to be 3 μm (Figure 6). A similarly made CFRP parabolic sample was smoothed using UV curable polymers and reflective layers deposited as described above to create the mirror that is demonstrated in this paper. An edge piece from the same finished mirror was characterized with an atomic force microscope (AFM) (Figure 7). The measured RMS surface roughness of the mirror was 4.5 nm, giving us 3 orders of magnitude improvement in surface roughness. Similar values of RMS roughness were measured before and after metallization.

For surfaces with subwavelength roughness, assuming Gaussian distribution of surface height, fraction of light scattered upon reflection is given by [8, 9] as

$$1 - \exp\left[-\left(\frac{4\pi\sigma\cos\theta}{\lambda}\right)^2\right]$$

where $\sigma$ is the RMS roughness, $\theta$ is the incidence angle, and $\lambda$ is the wavelength. If we desire to limit scattering loss to 2% at normal incidence, a mirror with 4.5 nm surface roughness is suitable for wavelengths above 400 nm.

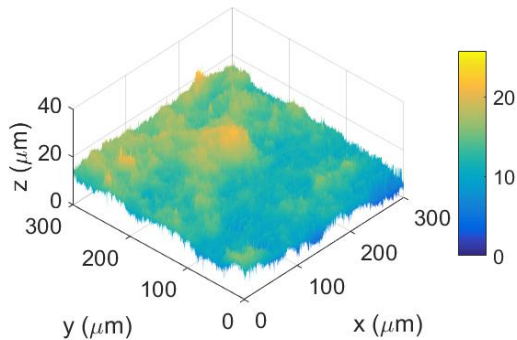

Fig. 6. Surface topography rendering of as-cast carbon fiber surface, measured by laser scanning confocal microscopy. The RMS surface roughness calculated from data is 3 μm.

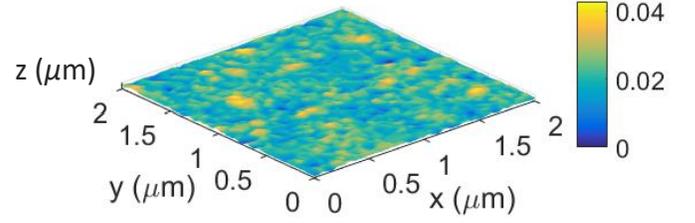

Fig. 7. Surface topography rendering of a finished mirror, measured by AFM. The RMS surface roughness is 4.5 nm.

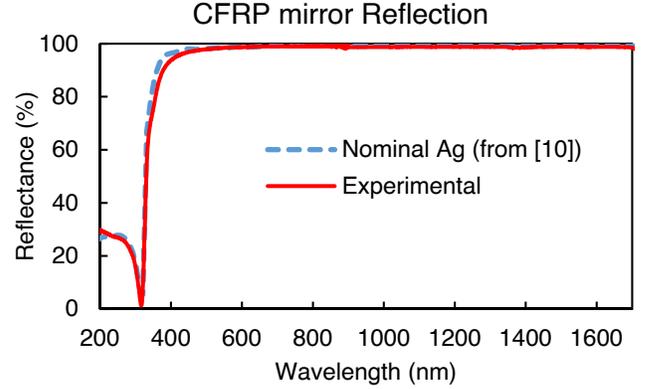

Fig. 8. Experimental reflectance data for a CFRP mirror, measured using a spectrophotometer. Also plotted is the nominal reflectance of polished Ag.

Figure 8 shows the experimental spectral reflectivity data for a CFRP reflector, taken using a Cary 5000 spectrophotometer, which agrees well with expectation for a polished Ag surface [10]. The roughness measurements and reflectance data confirm that our concentrator mirrors are adequately smooth for use over the solar spectrum down to wavelengths of about 400 nm, and that the primary limitation at shorter wavelengths is due to the Ag itself rather than surface scattering. Extending the usable range to ultraviolet wavelengths will require the use of a different reflective layer such as Al or dielectric-enhanced Ag, and may benefit from further reductions in surface roughness.

## V. PERFORMANCE VALIDATION

The optimal alignment was determined by mounting the concentrator on a translation stage under an AM0 solar simulator, and adjusting the distance to the receiver to maximize short circuit current ($I_{SC}$). We determined the optical efficiency, as defined by:

$$\eta = \left(\frac{I_{SC\_mirror}}{I_{SC\_cell}} \frac{1}{C}\right)$$

where $I_{SC\_mirror}$ is the short circuit current recorded from the cell with the parabolic mirror under illumination, and $I_{SC\_cell}$ is short circuit current recorded from the cell alone under illumination without the concentrator. $C$ is the geometric concentration ratio, defined as the ratio of the concentrator aperture area to the cell area, for a 2.5D concentrator this translates to a ratio of aperture width to cell width. The mirror has a nominal 15 mm optical width and the receiver cells used for this demonstration were 1.4 mm wide giving us a geometric concentration of 11x. Figure 9 shows a photograph of a concentrator pair during the alignment process. The reflected image of the cells (width of 1.4 mm) is enlarged and distributed across the entire focusing mirror (width of 15 mm), which indicates good optical performance. Figure 10 shows the measured optical efficiency of this concentrator pair at different aperture values. Here the peak optical efficiency of 77.5% corresponds to the point of zero position offset, which is the designed 15 mm aperture width. The peak optical efficiency was achieved at an incidence angle of 2 degrees, as previously predicted by ray tracing analysis of the same shape (Figure 5). The difference between the optical efficiency of 80% predicted by the ray trace analysis and the measured value can be accounted for by alignment errors and reflectance losses in the mirror coating.

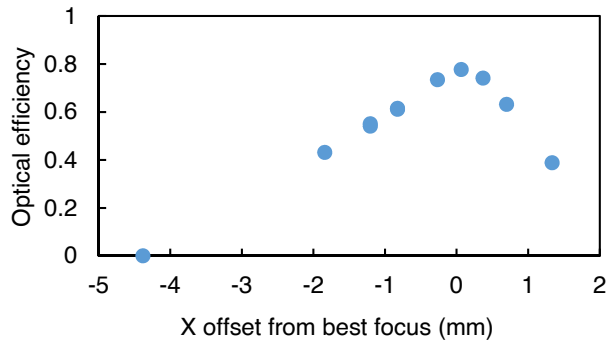

Fig. 10. CFRP mirror experimental optical efficiency curve. Peak optical efficiency is 77.5%

## VII. Conclusions

A proof-of-concept 11x concentrator made with lightweight carbon fiber parabolic mirrors achieved up to 77.5% optical efficiency. The primary loss of efficiency was due to shape deviation from the nominal parabola, as the smoothed surface of the mirrors provided excellent specular reflectance over the visible and near infrared wavelengths. Further optimization of this system will include improving the shape, reducing the thickness of the polymer smoothing layer, and making thinner CFRP composite for overall mass reduction. In addition, we will investigate the stability of the system in vacuum and under elevated temperatures and thermal cycling consistent with operation in space. Overall, this UV curable nano-meter scale smoothing process for making CFRP mirrors offers promising performance for an ultra-light concentrated photovoltaic system intended for space applications.

## Acknowledgements

We acknowledge funding from Northrop Grumman Corporation. This effort made use of facilities provided by the Kavli Nanoscience Institute, the Molecular Materials Research Center, the Resnick Institute, and the Joint Center for Artificial Photosynthesis at Caltech. We acknowledge the helpful contributions of Mark Kruer, Mike Levesque, and Erik Kurman at Northrop Grumman.

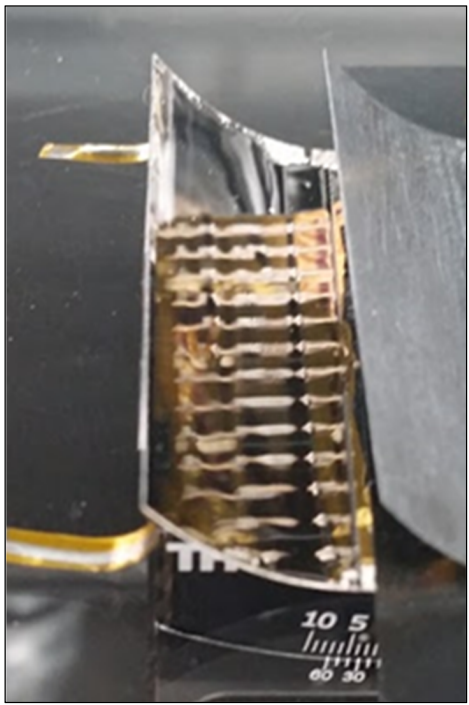

Fig. 9. Photograph of the image of the cells stretched clearly across the mirror indicating that the cells are positioned at the focus of this high quality mirror. Experimental configuration for testing the concentrator. The front CFRP parabolic shape supports a 1.4 mm cell at the top back edge.

REFERENCES

[1] P. Jaffe and J. McSpadden, "Energy Conversion and Transmission Modules for Space Solar Power," *Proceedings of the IEEE,* vol. 101, pp. 1424-1437, 2013.

[2] M. O'Neill, A. McDanal, M. Piszczor, P. George, M. Eskenazi, M. Botke*, et al.*, "Recent progress on the stretched lens array (SLA)," presented at the 18th Space Photovoltaic Research and Technology Conference, 2005.

[3] A. Rabl, "Optical and thermal properties of compound parabolic concentrators," *Solar Energy,* vol. 18, pp. 497-511, 1976.

[4] T. G. Stern, "Interim results of the SLATS concentrator experiment on LIPS-II (space vehicle power plants)," in *Conference Record of the Twentieth IEEE Photovoltaic Specialists Conference*, pp. 837-840 vol.2, 1988.

[5] N. Vaidya, T. E. Carver, and O. Solgaard, "Device fabrication using 3D printing," U.S. Provisional Patent Application No. 62/267,175, 2015.

[6] P. Espinet-Gonzalez, T. Vinogradova, M. D. Kelzenberg, A. Messer, E. Warmann, C. Peterson*, et al.*, "Impact of Space Radiation Environment on Concentrator Photovoltaic Systems," presented at the 44th IEEE PVSC, Washington DC, 2017.

[7] C. A. Silva, E. Marotta, M. Schuller, L. Peel, and M. O'Neill, "In-Plane Thermal Conductivity in Thin Carbon Fiber Composites," *Journal of Thermophysics and Heat Transfer,* vol. 21, pp. 460-467, 2007.

[8] J. B. Steeves, "Multilayer Active Shell Mirrors," Ph.D. Thesis, California Institute of Technology, Pasadena, CA, 2015.

[9] H. Davies, "The reflection of electromagnetic waves from a rough surface," *Proceedings of the IEE - Part IV: Institution Monographs,* vol. 101, pp. 209-214, 1954.

[10] P. B. Johnson and R. W. Christy, "Optical Constants of Noble Metals," *Physical Review B,* vol. 6, pp. 4370-4379, 1972.